\documentclass[12pt]{article}
\usepackage{epsf}
\usepackage{cite}
\usepackage{amsmath,amssymb}
\usepackage{color}

\usepackage[dvips]{graphicx}

\usepackage{comment}

\begin{document}

\newcommand{\lsim}{\stackrel{<}{_\sim}}
\newcommand{\gsim}{\stackrel{>}{_\sim}}

\setcounter{footnote}{0}

\begin{titlepage}

\def\thefootnote{\fnsymbol{footnote}}

\begin{center}

\hfill UT-14-35\\
\hfill August, 2014\\

\vskip .75in

{\large \bf 
Cosmological Implications of High-Energy Neutrino Emission 
\\
from the Decay of Long-Lived Particle
}

\vskip .75in

{\large 
Yohei Ema, Ryusuke Jinno and Takeo Moroi
}

\vskip 0.25in

\vskip 0.25in

{\em Department of Physics, University of Tokyo,
Tokyo 113-0033, Japan}

\end{center}
\vskip .5in

\begin{abstract}

  We study cosmological scenarios in which high-energy neutrinos are
  emitted from the decay of long-lived massive particles at the cosmic
  time later than a redshift of $\sim 10^6$.  The high-energy neutrino
  events recently observed by the IceCube experiment suggest a new
  source of high-energy cosmic-ray neutrinos; decay of a heavy
  particle can be one of the possibilities.  We calculate the spectrum
  of the high-energy neutrinos emitted from the decay of long-lived
  particles, taking account of the neutrino scattering processes with
  background neutrinos. Then, we derive bounds on the scenario using
  the observation of high-energy cosmic-ray neutrino flux.  We also
  study constraints from
  the spectral distortions of the cosmic
  microwave background and the big-bang nucleosynthesis.
  In addition, we show that the PeV neutrinos observed by the IceCube
  experiment can originate from the decay of a massive particle with
  its mass as large as $O(10^{10}\ {\rm GeV})$.

\end{abstract}

\end{titlepage}

\renewcommand{\thepage}{\arabic{page}}
\setcounter{page}{1}
\renewcommand{\thefootnote}{\#\arabic{footnote}}
\setcounter{footnote}{0}
\renewcommand{\theequation}{\thesection.\arabic{equation}}

\section{Introduction}
\label{sec:intro}
\setcounter{equation}{0}

In large classes of particle-physics models, there exist massive
long-lived particles.  Even though they may not be accessed by the
currently available colliders, information about those particles may
be obtained from astrophysical and cosmological observations.  If they
are produced in the early universe and also if they decay in or near
the present epoch, their decay products may affect the fluxes of
high-energy cosmic rays, resulting in constraints on their relic
densities, lifetimes, decay modes, and so on.  In addition, models
with long-lived particles have been attracted attentions to explain
the results of cosmic-ray observations \cite{Ibarra:2013cra}.  In
particular, implications of the decay processes into $\gamma$,
$e^\pm$, and (anti-) proton have been extensively studied.

With the successful detections of high-energy cosmic-ray neutrino
events at IceCube \cite{Aartsen:2013jdh,Aartsen:2014gkd}, our
understanding about the cosmic-ray neutrino flux is also significantly
improving.  In particular, the IceCube collaboration claims that the
cosmic-ray neutrino flux in the sub-PeV to PeV region is well above
that of expected backgrounds, which suggests a new source of
high-energy cosmic-ray neutrinos.  After the IceCube results are
released, it has been discussed that the decay of heavy particles may
be responsible for the IceCube events
\cite{Ema:2013nda,Feldstein:2013kka,Esmaili:2013gha,
  Bai:2013nga,Higaki:2014dwa,Bhattacharya:2014vwa,Zavala:2014dla,
  Chen:2014lla}.\footnote
{ For other explanations for the IceCube result, including
  astrophysical ones, see the review \cite{Anchordoqui:2013dnh} and
  the references therein.  }
Importantly, the lifetime of the long-lived particle (potentially)
responsible for the IceCube events can be either longer or shorter
than the present cosmic time.  In particular, the present authors have
argued that the decay of a long-lived particle (called $X$) in the
past can be the origin of the high-energy neutrinos observed by
IceCube \cite{Ema:2013nda}.  We call this scenario ``early-decay
scenario,'' in contrast to the ones with decaying dark matter. In the
previous study, we have calculated the neutrino flux originating from
the decay of $X$ for the case where the neutrino scattering processes
with background neutrinos are negligible (which is the case when the
decay of $X$ occurs at the epoch of $1+z\lsim 10^4$, with $z$ being
the redshift), and have pointed out that the IceCube events may be
well explained in this scenario.

In this paper, we extend our previous study and discuss astrophysical
and cosmological constraints on the early-decay scenario.  We pay
particular attention to the effects of the neutrino scattering
processes with background neutrinos, which are not completely taken
into account in our previous study. We calculate the flux of
cosmic-ray neutrinos originating from the decay of $X$.  Then,
comparing the result with the observed cosmic-ray neutrino flux, we
derive an upper bound on the primordial abundance of $X$.  In
addition, photons and charged particles are also produced in
association with the neutrino scattering processes; they result in the
spectral distortion of the cosmic microwave background (CMB) and the
change of the light-element abundances produced by the big-bang
nucleosynthesis (BBN), from which we obtain an upper bound on the
abundance of $X$.  For the study of the constraints from the CMB
distortion, we take into account the present bound (COBE/FIRAS
\cite{Fixsen:1993rd,Fixsen:1996nj}), or the expected bound in the
future (for example, PIXIE \cite{Kogut:2011xw} and PRISM
\cite{Andre:2013nfa}).  In our study, we will not specify the detailed
particle-physics model which contains the candidate of $X$, but we
perform our analysis as general as possible.  We discuss the
constraints on the scenario using the properties of $X$, i.e., its
lifetime, energy distribution of the final-state neutrinos (which is
assumed to be monochromatic in this paper), and its primordial relic
density.  We also discuss the implication of the IceCube result in
light of the early-decay scenario.  In particular, it may be possible
that the neutrino excess in sub-PeV region and the possible cutoff
around PeV are simultaneously explained if we take into account the
effect of the neutrinos scattered by the background neutrinos.

The organization of this paper is as follows.  In
Sec.~\ref{sec:effects} we discuss the evolution of the neutrino flux
originating from the massive decaying particle $X$.  The effects of
the produced neutrinos on the CMB distortions and the light-element
abundances are also explained there.  Then, in the following section,
we give constraints on the primordial relic density of $X$ using the
observation of the cosmic-ray neutrino flux, the CMB distortions and
the light-element abundances produced by the BBN.  In
Sec.~\ref{sec:implication} we discuss possible interpretations of the
recent IceCube high-energy neutrino events in our scenario. The final
section is devoted to the conclusions and discussion.

\section{Effects of Neutrino Emission}
\label{sec:effects}
\setcounter{equation}{0}

\subsection{Evolution of Neutrino Flux}
\label{sec:evol}

Let us first discuss the evolution of the neutrino flux produced by
the decay of the parent particle $X$.  Once produced, the neutrinos
propagate in the expanding universe scattering off background
particles (in particular, neutrinos).  Then, in order to obtain the
neutrino flux in $l$-th flavor, $\Phi_{\nu,l}(t, E)$, which is related
to the number density of the $l$-th flavor neutrino as $n_{\nu, l}(t)
= \int dE \Phi_{\nu, l}(t, E)$, we solve the following Boltzmann
equation:
\begin{align}
  \left(\frac{\partial}{\partial t} + 2 H
    - H E \frac{\partial}{\partial E} \right) \Phi_{\nu, l}(t, E) 	
  = &
  -\gamma_{\nu, l}(t;E) \Phi_{\nu, l}(t, E) 
  \nonumber \\ & +
  \int dE^{\prime} 
  \Phi_{\nu, n}(t, E^{\prime})
  \frac{d \gamma_{\nu, nm}(t;E^{\prime}, E)}{d E}
  P_{ml}(t,E) \nonumber \\ &
  + S_{\nu, m}(t, E) P_{ml}(t,E),
  \label{eq:Boltzmann}
\end{align}
where $H$ is the expansion rate of the universe, and $S_{\nu, l}(t,
E)$ is the source term.  In addition $\gamma_{\nu, l}(t; E)$ is the
scattering rate, and $d\gamma_{\nu, ml}(t;E^{\prime}, E)/d E$ is the
(differential) neutrino production 
 rate with $E^{\prime}$ and $E$
being the energies of initial- and final-state neutrinos.  (Here, $m$
and $n$ are flavor indices; summation over these indices is
implicit.)  At the cosmic time when the neutrino scattering processes
become effective, at which the scattering rate becomes important,
$\gamma_{\nu, l}(t; E)$ is (almost) flavor-independent.  Thus, we take
$\gamma_{\nu, l}(t; E) = \gamma_{\nu}(t; E)$.
In our calculation, the effect of the neutrino oscillation is taken
into account by introducing the ``transition probability'' $P_{ml}(t,
E)$.  We approximate that the flavors are fully mixed in the case
where the time scale of the neutrino oscillation (i.e., $2E/|\Delta
m_{21}|^{2}$ or $2E/|\Delta m_{31}|^{2}$) is shorter than the mean free
time (i.e., $\gamma_{\nu}^{-1}$), and that the effect of the neutrino
oscillation is negligible in the opposite case.  (Here, $|\Delta
m_{21}|^{2}$ and $|\Delta m_{31}|^{2}$ are the neutrino mass squared
differences.)  Then, taking $|\Delta m_{21}|^{2} = 7.50\times 10^{-5}\
{\rm eV}^2 $, $|\Delta m_{31}|^{2} = 2.47\times 10^{-3}\ {\rm eV}^2$,
$\sin^2 \theta_{12} = 0.30$, $\sin^2 \theta_{13} = 0.023$, and $\sin^2
\theta_{23} = 0.41$ \cite{GonzalezGarcia:2012sz} with $\theta$'s
being the mixing angles in the neutrino mixing matrix,\footnote
{In our approximation, $P_{ml}(t, E)$ is evaluated at the time of the
  neutrino emission.  Therefore, if a sizable amount of neutrino
  propagated from the epochs of 1 or 2 to the present epoch without
  being scattered, we would fail to include the effects of neutrino
  oscillation during the propagation.  However in reality,
  $\gamma_{\nu}(t; E) < |\Delta m_{21}|^2/2E$ for $\tau(t, E/(1+z(t)))
  \lsim 10$, where $\tau(t, E/(1+z(t)))$ is the optical depth of
  neutrino defined in Eq.\ (\ref{eq:optical depth}), and such a problem
  does not occur.}
$P_{ml}(t, E)$ is evaluated as follows:
\begin{itemize}
\item[1.] When $|\Delta m_{31}|^2/2E < \gamma_{\nu}(t; E)$, the
  scattering time scale is shorter than those of neutrino
  oscillation. In this case, the effect of neutrino oscillation is
  neglected and we take $P_{ml}(t, E) = {\rm diag}(1, 1, 1)$.
\item[2.] When $|\Delta m_{21}|^2/2E < \gamma_{\nu}(t; E) < |\Delta
  m_{31}|^2/2E$, the neutrino oscillation due to $\Delta m_{21}$ is
  neglected, while the oscillation due to $\Delta m_{31}$ is taken
  into account. In this case, we take:
  \begin{align}
    P_{e\mu}(t, E) &= 0.02, \nonumber \\
    P_{e\tau}(t, E) &= 0.03, \nonumber \\
    P_{\mu\tau}(t, E) &= 0.47.
  \end{align}
\item[3.] When $\gamma_{\nu}(t; E) < |\Delta m_{21}|^2/2E$, we
  approximate that neutrino oscillations due to $\Delta m_{21}$ and
  $\Delta m_{31}$ are so fast that the full mixing of the neutrino
  flavors is realized. In this case, we take:
  \begin{align}
    P_{e\mu}(t, E) &= 0.28, \nonumber \\
    P_{e\tau}(t, E) &= 0.16, \nonumber \\
    P_{\mu\tau}(t, E) &= 0.37.
  \end{align}
\end{itemize}
Here, we neglect the $CP$-violation in the neutrino mixing, and hence
we take $P_{ml}(t, E) = P_{lm}(t, E)$.  The diagonal
elements of $P_{ml}(t, E)$ can be evaluated by using $\sum_m P_{lm}(t,
E)=1$.

In order to take into account the effects of neutrino scattering, we
consider the following scattering processes with background (anti-)
neutrinos:\footnote
{If the energy of the injected neutrino is very high, the scattering
  process with background photons also becomes relevant.  However, we
  have checked that the following results do not change even if we
  take such a process into account.}
\begin{itemize}
\item $\nu_{l}+\nu_{l,\text{BG}} \rightarrow \nu_{l} + \nu_{l}$,
\item $\nu_{l}+\nu_{l^{\prime},\text{BG}} \rightarrow \nu_{l} + \nu_{l^{\prime}}$,
  with $l\neq l^{\prime}$,
\item $\nu_{l}+\bar{\nu}_{l,\text{BG}} \rightarrow \nu_{l} + \bar{\nu}_{l}$,
\item $\nu_{l}+\bar{\nu}_{l,\text{BG}} \rightarrow l + \bar{l}$,
\item $\nu_{l}+\bar{\nu}_{l,\text{BG}} \rightarrow f + \bar{f}$,
  with $f \neq l, \nu_{l}$,
\item $\nu_{l}+\bar{\nu}_{l^{\prime},\text{BG}} \rightarrow \nu_{l} +
  \bar{\nu}_{l^{\prime}}$, with $l\neq l^{\prime}$,
\item $\nu_{l}+\bar{\nu}_{l^{\prime},\text{BG}} \rightarrow l +
  \bar{l}^{\prime}$, with $l\neq l^{\prime}$,
\end{itemize}
where $l=e$, $\mu$, $\tau$, while $f$ denotes the standard-model
fermions, and the subscript ``BG'' is used for background neutrinos.
The scattering rate $\gamma_{\nu}(t; E)$ is calculated with taking
into account the effects of these processes.

In the calculation of the neutrino production rate, we include two
contributions as
\begin{align}
  \frac{d\gamma_{\nu, ml}}{d E} = 
  \frac{d\gamma^{\rm (dir)}_{\nu, ml}}{d E} + 
  \frac{d\gamma^{(\gamma\gamma)}_{\nu, ml}}{d E}.
  \label{dgamma/dE}
\end{align}
One is the neutrinos directly produced by the scattering processes
listed above, which corresponds to the first term of the right-hand
side of Eq.\ \eqref{dgamma/dE}. In the neutrino-neutrino scattering
processes, energetic neutrinos are produced directly or by the decay
of final-state particles.  (Notice that the standard-model fermions
other than neutrinos and $e^\pm$ undergo hadronization and/or decay
processes after the production.)  In our numerical calculation, we
calculate the energy distributions of the neutrinos (as well as other
stable particles, i.e, $e^\pm$, $\gamma$, $p$ and $\bar{p}$) produced
by the scattering processes listed above using PYTHIA package
\cite{Sjostrand:2006za, Sjostrand:2007gs}.  The other is the neutrinos
produced by double-photon pair creations of standard-model fermions
(the second term of the right-hand side of Eq.\ \eqref{dgamma/dE}).  A
sizable amount of high-energy photons may be produced as a consequence
of neutrino-neutrino scattering processes (after the hadronization
and/or decay of colored particles).  In addition, high-energy $e^\pm$s
produced by the neutrino scattering processes are converted to
high-energy photons via the inverse Compton process.  By scattering
off the CMB, those high-energy photons may induce double-photon pair
creations of standard-model fermions whose decay products contain
neutrinos.  Then, we estimate
\begin{align}
  \frac{d\gamma^{(\gamma\gamma)}_{\nu, ml}(E',E)}{d E}
  = \int d\epsilon
  \left( 
    \frac{d\gamma_{\gamma,m}(E',\epsilon)}{d \epsilon}
    + \frac{d\gamma_{e^\pm,m}(E',\epsilon)}{d \epsilon}
  \right) 
  \frac{dN^{(\gamma\gamma)}_{\nu, l}(\epsilon,E)}{d E},
\end{align}
where $d\gamma_{\gamma,m}/d \epsilon$ and $d\gamma_{e^\pm,m}/d
\epsilon$ are (differential) production rate of photon and $e^\pm$ via
the neutrino scattering processes, respectively.  (Here, $\epsilon$
denotes the energy of $\gamma$ or $e^\pm$ produced by the neutrino
scattering processes.)  We approximate that the energy of the photon
produced by the inverse Compton scattering is equal to that of the
initial-state $e^\pm$.  This is because, in the center-of-mass frame,
the inverse Compton scattering is significantly enhanced for backward
scattering in the relativistic limit \cite{Berestetsky:1982aq}.  In
addition, $dN^{(\gamma\gamma)}_{\nu,l}/dE$ is the spectrum of
neutrinos (after the hadronization and/or the decay processes)
produced as a consequence of the (multiple) double-photon pair
creation.  Using the fact that, in the center-of-mass frame, the
double-photon pair creation cross section is sharply peaked when the
momenta of final-state fermions are parallel (or anti-parallel) to
those of initial-state photons \cite{Berestetsky:1982aq}, we
approximate that the energy of one of the final-state fermions is
equal to that of initial-state high-energy photon while that of
another fermion is negligibly small.  Thus, with the injections of
photon and $e^\pm$, electromagnetic cascade occurs.  During the
cascade, the energy of the electromagnetic sector is reduced via the
emission of neutrino, which is due to the decay of unstable particles
like muon.  We approximate that the double-photon pair productions of
the fermions other than $e^\pm$ become ineffective for the photon with
the energy $E$ once the ratio of the scattering rates
$\Gamma_{\gamma\gamma\rightarrow\mu^+\mu^-}/\Gamma_{\gamma\gamma\rightarrow
  e^+e^-}$ becomes smaller than $m_e^2/ET$ (with $m_e$ being the
electron mass); here, we use the fact that $E_{\rm i}-E_{\rm f}\sim
O(m_e^2/T)$, where $E_{\rm i}$ and $E_{\rm f}$ are energies of
energetic initial- and final-state particles in the processes
$\gamma\gamma\rightarrow e^+e^-$ and $e^\pm\gamma\rightarrow
e^\pm\gamma$.  The remaining electromagnetic particles may also affect
the CMB spectrum and the light-element abundances, as we will discuss 
in Sec.\ \ref{sec:distortion} and \ref{sec:BBN}, respectively.

When the neutrinos are produced by the decay of $X$, the source term
is given by
\begin{align}
  S_{\nu, l}(t, E) = 
  \frac{1}{4\pi}\frac{n_{X}(t)}{\tau_{X}}\frac{d N_{\nu, l}^{(X)}}{dE},
\end{align}
where $n_{X}(t)$ is the number density of $X$, $\tau_{X}$ is the
lifetime of $X$, and $dN_{\nu, l}^{(X)}/dE$ is the energy distribution
of the $l$-th flavor neutrinos produced by the decay of $X$.  Using
the so-called yield variable $Y_{X}$ defined as
\begin{align}
  Y_{X} \equiv \left[\frac{n_{X}(t)}{s(t)} \right]_{t \ll \tau_{X}},
\end{align}
with $s(t)$ being the entropy density, $n_{X}(t)$ is given by
\begin{align}
  n_{X}(t) = Y_{X} s(t) e^{-t/\tau_{X}}.
\end{align}
For simplicity, we consider only the case where the neutrinos produced
by the decay of $X$ are monochromatic (with the energy of
$\bar{E}_\nu$). Then, $dN_{\nu, l}^{(X)}/dE$ is given by
\begin{align}
  \frac{d N_{\nu, l}^{(X)}}{dE} = \bar{N}_{\nu, l}\delta(E - \bar{E}_{\nu}),
\end{align}
where $\bar{N}_{\nu, l}$ is the number of $l$-th flavor neutrinos
produced by the decay of one $X$.  For simplicity, we consider the
case where the decay of $X$ produces equal amount of $e$, $\mu$, and
$\tau$ neutrinos, taking $\bar{N}_{\nu, e} = \bar{N}_{\nu, \mu} =
\bar{N}_{\nu, \tau} = 1/3$.\footnote
{We have checked that, because of the neutrino oscillation, the
  resultant neutrino flux for each flavor does not depend much on this
  assumption. }
Moreover, we also assume that the $CP$
violation is negligible in the decay of $X$ and that the fluxes of
neutrinos and anti-neutrinos are equal.  
We note that particles other than neutrinos such as electrons and
photons may also be produced by the decay of $X$. The cosmological
implications of such particles are discussed, for example, in
\cite{Hu:1993gc, Furlanetto:2006wp, Valdes:2007cu, Chluba:2011hw,
  Chen:2003gz, Zhang:2007zzh, Slatyer:2012yq, Khatri:2012tw, Chluba:2013vsa,
  Chluba:2013wsa, Chluba:2013pya}.  Thus, in our analysis,
the present neutrino flux is determined by the following three
parameters:
\begin{align}
  \bar{E}_{\nu},\  z_{*} \equiv z(\tau_{X}),\  Y_{X}.
\end{align}

\begin{figure}[t]
  \begin{center}
    \includegraphics[width = 9cm]{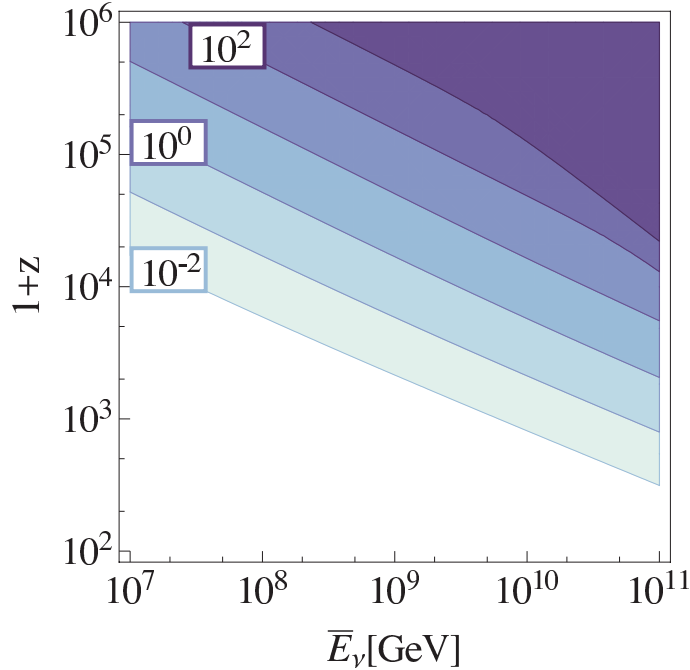}
    \caption{\small Contours of constant $\tau(z;E)$, with $E$ being
      the present energy of the neutrino.  The contours are
      $\tau=0.01$, $0.1$, $\cdots$, and $1000$, from bottom to top.
      (The numbers in the figure give the value of $\tau$.)  The
      horizontal axis is the initial energy of the neutrino
      $\bar{E}_\nu=(1+z)E$, while the vertical axis is the redshift
      $1+z$.}
    \label{fig:optical_depth}
  \end{center}
\end{figure}

The neutrino flux at the present cosmic time $t_0$ 
can be decomposed into two contributions:
\begin{align}
  \Phi_{\nu, l}(t_0, E) = 
  \Phi_{\nu, l}^{(\text{prim})}(E) + \Phi_{\nu, l}^{(\text{sec})}(E).
  \label{prim+sec}
\end{align}
Here, $\Phi_{\nu, l}^{(\text{prim})}(E)$ is the flux of neutrinos
which propagate to the present epoch without being scattered (which we
call ``primary neutrinos''), while $\Phi_{\nu, l}^{(\text{sec})}(E)$
is that of secondary neutrinos produced by the neutrino scattering
processes and the electromagnetic cascade.  With the monochromatic
neutrino injection, $\Phi_{\nu, l}^{(\text{prim})}(E)$ is given
by\footnote{ The effects of neutrino oscillation do not appear in Eq.\
  (\ref{eq:flux}) because we assume that the neutrinos produced by the
  decay of $X$ are flavor-universal.  }
\begin{align}
  \Phi_{\nu, l}^{(\text{prim})}(E) = 
  \frac{1}{4\pi}\frac{\bar{N}_{\nu, l}Y_{X}s(t_{0})}{\tau_{X}E}
  \left[
    \frac{e^{-\bar{t}/\tau_{X}} 
      e^{-\tau(z(\bar{t}); E)}}{H(\bar{t})}
  \right]_{1+z(\bar{t})=\bar{E}_{\nu}/E},
  \label{eq:flux}
\end{align}
where $\tau(z;E)$ is the optical depth of neutrinos
\begin{align}
  \tau(z; E) \equiv \int_{t(z)}^{t_{0}} dt^{\prime}
  \gamma_{\nu}(t^{\prime}; (1+z(t^{\prime})) E).
  \label{eq:optical depth}
\end{align}
From Eq.\ (\ref{eq:flux}), one can see that the neutrinos produced at
higher redshifts contribute to the present flux at lower energies.

In order to see when the neutrino scattering is effective, in Fig.\
\ref{fig:optical_depth}, we plot the contours of constant $\tau(z;E)$
on $\bar{E}_\nu\equiv (1+z)E$ vs.\ $1+z$ plane.  As one can see, the
optical depth increases as $\bar{E}_\nu$ or $z$ becomes larger; this
is because the neutrino scattering cross section is more enhanced with
higher center-of-mass energy $E_{\rm CM}$ (as far as $E_{\rm CM}<
m_Z$).  The optical depth becomes $\sim 1$ when the scattering rate of
the neutrino is comparable to the expansion rate of the universe.  For
the present energy of $E=10^6$ and $10^7\ {\rm GeV}$, for example,
$\tau(z;E)\gtrsim 1$ is realized when $1+z\gtrsim 10^{4}$ and $3\times
10^{3}$, respectively.  When $\tau \gsim 1$, the neutrino scattering
processes become important and $\Phi_{\nu, l}^{(\text{sec})}(E)$ is
sizable. One can also see that some of the contours show bending
behavior.  This is due to the change in the neutrino cross section at
$E_{\rm CM} \sim m_Z$.

We numerically evaluate the neutrino flux by solving Eq.\
\eqref{eq:Boltzmann}.  For this purpose, we introduce the Green's
function $G_{ml}(t^{\prime}, E^{\prime}; t, E)$, which satisfies
\begin{align}
  \left( \frac{\partial}{\partial t} + 2 H
    - H E \frac{\partial}{\partial E} \right)
  G_{ml}(t^{\prime}, E^{\prime}; t, E) = 
  &- \gamma_{\nu}(t; E)G_{ml}(t^{\prime}, E^{\prime}; t, E)
  \nonumber \\ &+
  \int dE^{\prime\prime} G_{mn}(t^{\prime}, E^{\prime}; t, E^{\prime\prime}) 
  \frac{d \gamma_{\nu, np}(t; E^{\prime\prime}, E)}{dE} P_{pl}(t, E)
  \nonumber \\ &
  + \delta_{ml}\delta(t^{\prime} - t)\delta(E^{\prime} - E),
\end{align}
and
\begin{align}
  G_{ml}(t^{\prime}, E^{\prime}; t, E)_{t^{\prime} > t} = 0.
\end{align}
With the Green's function, the neutrino flux is given by
\begin{align}
  \Phi_{\nu, l}(t, E) = \int_{0}^{t} dt^{\prime} \int dE^{\prime} 
  S_{m}(t^{\prime}, E^{\prime}) P_{mn}(t^{\prime}, E^{\prime})
  G_{nl} (t^{\prime}, E^{\prime}; t, E).
\end{align}
In order to evaluate $G_{ml}(t^{\prime}, E^{\prime}; t, E)$, we use
the fact that $G_{ml}(t^{\prime}, E^{\prime};t, E)$ satisfies the
following relation:
\begin{align}
  G_{ml}&(t^{\prime}, E^{\prime}; t, E) 
  = \theta(t - t^{\prime})
  \Bigg[ \left(\frac{a(t^{\prime})}{a(t)} \right)^2
  e^{- \tilde{\tau}(t^{\prime}, t;E^{\prime})}
  \delta_{ml}\delta(E^{\prime} - a(t)E/a(t^{\prime}))
  \nonumber \\
  +& 
    \int_{t^{\prime}}^{\infty}dt^{\prime\prime} \int dE^{\prime\prime}
    \left(\frac{a(t^{\prime})}{a(t^{\prime\prime})}\right)^3
    e^{-\tilde{\tau}(t^{\prime}, t^{\prime\prime}; E^{\prime})}
    \frac{d \gamma_{\nu, mn}(t^{\prime\prime};
      a(t^{\prime})E^{\prime}/a(t^{\prime\prime}), E^{\prime\prime})}
    {dE^{\prime\prime}}
    P_{np}(t^{\prime\prime}, E^{\prime\prime})
    G_{pl}(t^{\prime\prime}, E^{\prime\prime};t, E)
  \Bigg],
  \label{recursive}
\end{align}
where $a(t)$ is the scale factor at the cosmic time $t$, and
\begin{align}
  \tilde{\tau}(t^{\prime}, t; E^{\prime}) \equiv
  \int_{t^{\prime}}^t dt^{\prime\prime} 
  \gamma_{\nu}(t^{\prime\prime}; a(t^{\prime})E^{\prime}/a(t^{\prime\prime})).
\end{align}
In our numerical calculation, we discretize Eq.\ \eqref{recursive} and
recursively evaluate the Green's function, with which the neutrino
flux is calculated. When the neutrinos produced by the decay of $X$
are flavor-universal and neutrino oscillation is taken into account,
the present neutrino fluxes are almost flavor-universal.  Therefore,
in the following discussion, we neglect the flavor dependence of the
neutrino flux and use $\Phi_{\nu}(E)$, which is defined as
\begin{align}
  \Phi_{\nu}(E) \equiv 
  \frac{1}{3}
  \left[ 
    \Phi_{\nu, e}(t_{0}, E) + \Phi_{\nu,\mu}(t_{0}, E) +
    \Phi_{\nu, \tau}(t_{0}, E)
  \right].
\end{align}

\begin{figure}[t]
\begin{center}
  \includegraphics[width=0.95\textwidth]{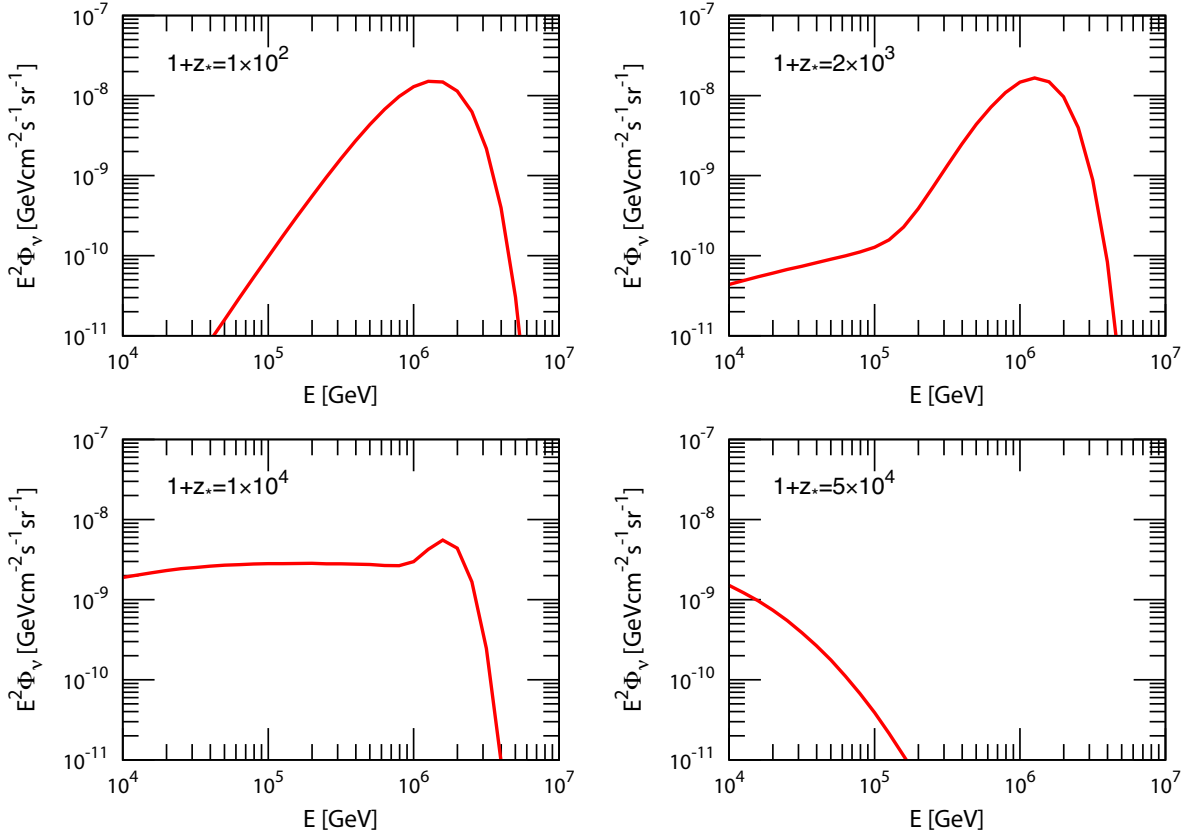}
  \caption{\small The present per-flavor neutrino fluxes for some
    different input parameters. Top left: $(\bar{E}_{\nu}, 1+z_{*},
    Y_{X}) = (10^{8}\ \text{GeV}, 10^{2}, 10^{-26})$. Top right:
    $(\bar{E}_{\nu}, 1+z_{*}, Y_{X}) = (2\times10^{9}\ \text{GeV},
    2\times10^{3}, 10^{-26})$. Bottom left: $(\bar{E}_{\nu}, 1+z_{*},
    Y_{X}) = (10^{10}\ \text{GeV}, 10^{4}, 10^{-26})$. Bottom right:
    $(\bar{E}_{\nu}, 1+z_{*}, Y_{X}) = (5\times10^{10}\ \text{GeV},
    5\times10^{4}, 10^{-26})$.}
  	\label{fig:present_flux}
\end{center}
\end{figure}

In Fig.\ \ref{fig:present_flux}, we show the neutrino fluxes at the
present epoch for several values of $z_{*}$.  Here, we take $Y_{X} =
10^{-26}$ and $\bar{E}_{\nu}/(1+z_{*}) = 1\ \text{PeV}$.  The
qualitative behavior of the neutrino spectrum can be understood as
follows: 
\begin{itemize}
\item[1.] With small enough $z_*$, the neutrino scattering is
  inefficient.  In such a case, the neutrino spectrum is affected only
  by the redshift and has a peak at $E \sim \bar{E}_{\nu}/(1+z_{*})$. 
  (See the top-left panel with $1 +
  z_{*} = 10^{2}$.)  
\item[2.] With the increase of $z_*$, a tail-like structure shows up 
because of the neutrino scattering processes.
  (See the top-right panel with $1 + z_{*} =2\times10^{3}$.) Comparing with the
  top-left panel, we can also see that the flux is slightly reduced
  at $2\times10^{5}\ \text{GeV} \lsim E \lsim 5\times10^{5}\
  \text{GeV}$ in the top-right panel.  
  This is due to the fact that
  the neutrinos with lower present energies are more likely to be affected 
  by the scattering processes with the background neutrinos because 
  they are produced at higher
  redshifts.
  In the top-right panel, the neutrino
  scattering thus reduces the flux at $2\times10^{5}\
  \text{GeV} \lsim E \lsim 5\times10^{5}\ \text{GeV}$, while the
  secondary neutrinos produced by the neutrino scattering 
  and the electromagnetic cascade contribute
  to the flux at $E \lsim 2\times10^{5}\ \text{GeV}$.
\item[3.] With larger $z_*$, $\bar{E}_{\nu}$ and $z_{*}$ become so
  large that a sizable fraction of neutrinos experience the
  scatterings with background neutrinos.  Consequently, the neutrino
  flux at around the peak is also reduced.  (See the bottom-left panel
  with $1+z_{*} = 10^{4}$.)
\item[4.] Then, with larger enough $z_*$, the neutrino scattering
  processes are so efficient that almost all the neutrinos emitted from 
  the $X$ decay are scattered.  (See the bottom-right panel with
  $1+z_{*} = 5\times10^{4}$.)  The present neutrino flux originates
  from secondary neutrinos.
\end{itemize}

\subsection{CMB Spectral Distortions}
\label{sec:distortion}

Next, we discuss the effects of electromagnetic particles produced by
the neutrino scattering processes.  In general, if photons or charged
particles are injected in the early universe, these particles may
affect the spectrum of the CMB. The type of the spectral distortion
relevant for the present scenario depends on the epoch at which the
energy injection occurs \cite{Zeldovich:1969ff, Sunyaev:1970er,
  Danese:1982, Burigana:1991, Hu:1992dc, Burigana:1995}:\footnote
{In our analysis, we approximate that the distorted spectrum can
    be parametrized by $y$ and $\mu$.  However, for $1.5\times10^{4}
    \lsim z \lsim 2\times10^{5}$, we might better consider
    intermediate-type distortions; for such an  analysis of the
    CMB distortion, see \cite{Chluba:2011hw, Khatri:2012tw,
      Chluba:2013vsa, Chluba:2013wsa, Chluba:2013pya}.  In addition,
    with such a precise analysis, we may have a chance to acquire
    information about the lifetime of $X$
    \cite{Chluba:2011hw, Chluba:2013vsa, Chluba:2013wsa, Chluba:2013pya}. }
\begin{itemize}
\item For $z\gsim 2 \times 10^6$, the complete thermalization is
  achieved and there is no spectral distortion.
\item For $5 \times 10^4 \lsim z \lsim 2 \times 10^6$, 
the kinetic equilibrium is realized while the chemical equilibrium is not. 
As a result, the so-called $\mu$-type distortion is produced.
\item For $z_{\rm rec} \lsim z \lsim 5 \times 10^4$, where $z_{\rm rec}$ is the 
redshift at the recombination, even the kinetic equilibrium is not
  achieved. Then, the so-called $y$-type distortion occurs.
\end{itemize}

In the case of the $\mu$-type distortion, 
the distribution
function of the CMB photon $f_\gamma(\omega)$ (with $\omega$ being the
energy of $\gamma$) becomes the Bose-Einstein distribution with the
chemical potential $\mu$ \cite{Hu:1992dc}: 
\begin{align}
  f_{\gamma} (\omega) = 
  \left[\exp\left(\frac{\omega}{T} + \mu\right)-1 \right]^{-1}.
\end{align}
Here, $T$ is the CMB temperature after the completion of the decay of
$X$ \cite{Chluba:2012gq}. The chemical potential is given by
\begin{align}
  \mu \simeq 1.4
  \int_{z_{K}}^{\infty} dz
  \frac{Q(z)}{\rho_{\text{rad}}(z)}J_{\mu}(z),
\end{align}
where $1+z_{K} = 5\times10^{4}$, $\rho_{\text{rad}}(z)$ is the radiation energy
density at the redshift $1+z$, and $Q(z)$ is the energy injection rate.
In addition, $J_{\mu}$ is the so-called distortion visibility function:
\begin{align}
  J_{\mu}(z) = \exp 
  \left[-\left( \frac{z}{2\times10^{6}} \right)^{2.5} \right],
\end{align}
which parametrize the fraction of injected energy at the redshift
$1+z$ converted into the $\mu$-type distortion.

In the case of the $y$-type distortion, the deviation of the CMB
spectrum from the black-body distribution is parametrized as
\cite{Zeldovich:1969ff}
\begin{align}
  \frac{\delta f_{\gamma}(\omega)}{f_{\gamma}(\omega)} = 
  y \frac{xe^{x}}{e^{x}-1}
  \left[x\left(\frac{e^{x} + 1}{e^{x} - 1}\right) - 4\right],
\end{align}
where $x \equiv \omega/T$, and the $y$ parameter is estimated as\footnote{
We set the lower bound of the integral in Eq.~(\ref{eq:y}) to $0$, 
since changing it below $z_{\rm rec}$ does not affect our result.
}
\begin{align}
  y \simeq 
  \frac{1}{4}\int_{0}^{z_{K}}dz \frac{Q(z)}{\rho_{\text{rad}}(z)}.
  \label{eq:y}
\end{align}
In the present scenario, $Q(z)$ comes from secondary photons and
charged particles produced by the neutrino scattering processes.  As
we have mentioned, we calculate the energy spectra of the stable
electromagnetic particles (i.e., $\gamma$, $e^\pm$, $p$ and $\bar{p}$)
using PYTHIA in order to evaluate $Q(z)$.  We assume that the
secondary photons and charged particles are instantaneously converted
into the $y$ or $\mu$ parameters after the double-photon pair
productions of the fermions other than $e^{\pm}$ become ineffective.
This is a good approximation in the case of our interest because the
interactions of photons and charged particles are fast enough when the
neutrino scattering is effective\cite{Slatyer:2009yq}.

\begin{figure}
  \begin{center}
    \includegraphics[width = 9cm]{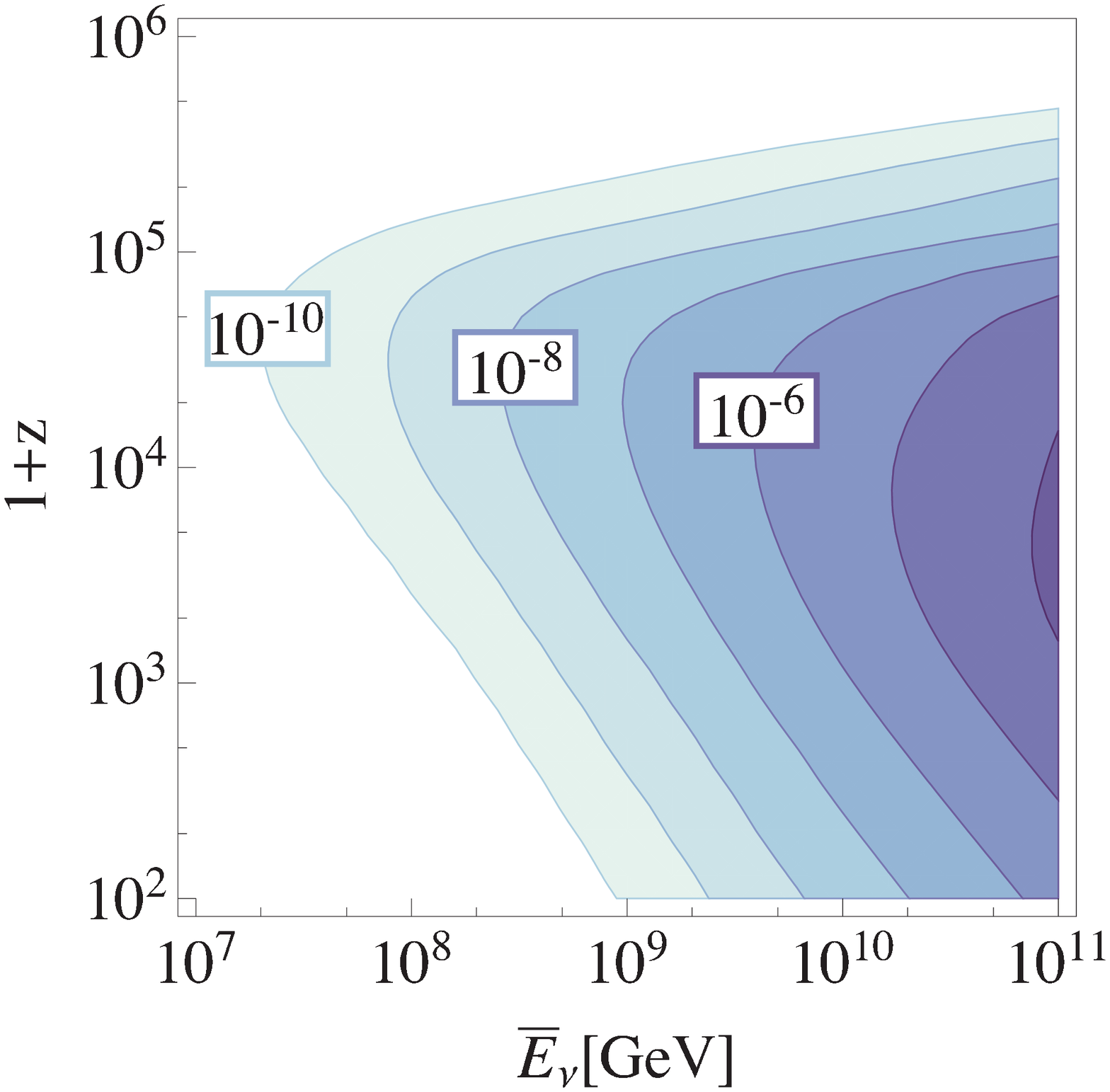}
    \caption{\small Contours of constant $y$ on $\bar{E}_\nu$ vs.\
      $1+z_*$ plane.  The contours are $y=10^{-10}$, $10^{-9}$,
      $\cdots$, and $10^{-4}$, from left to right.  (The numbers in the
      figure give the value of $y$.)  Here, we take $Y_{X} =
      10^{-22}$.}
    \label{fig:y-param}
    \vspace{10mm}
       \includegraphics[width = 9cm]{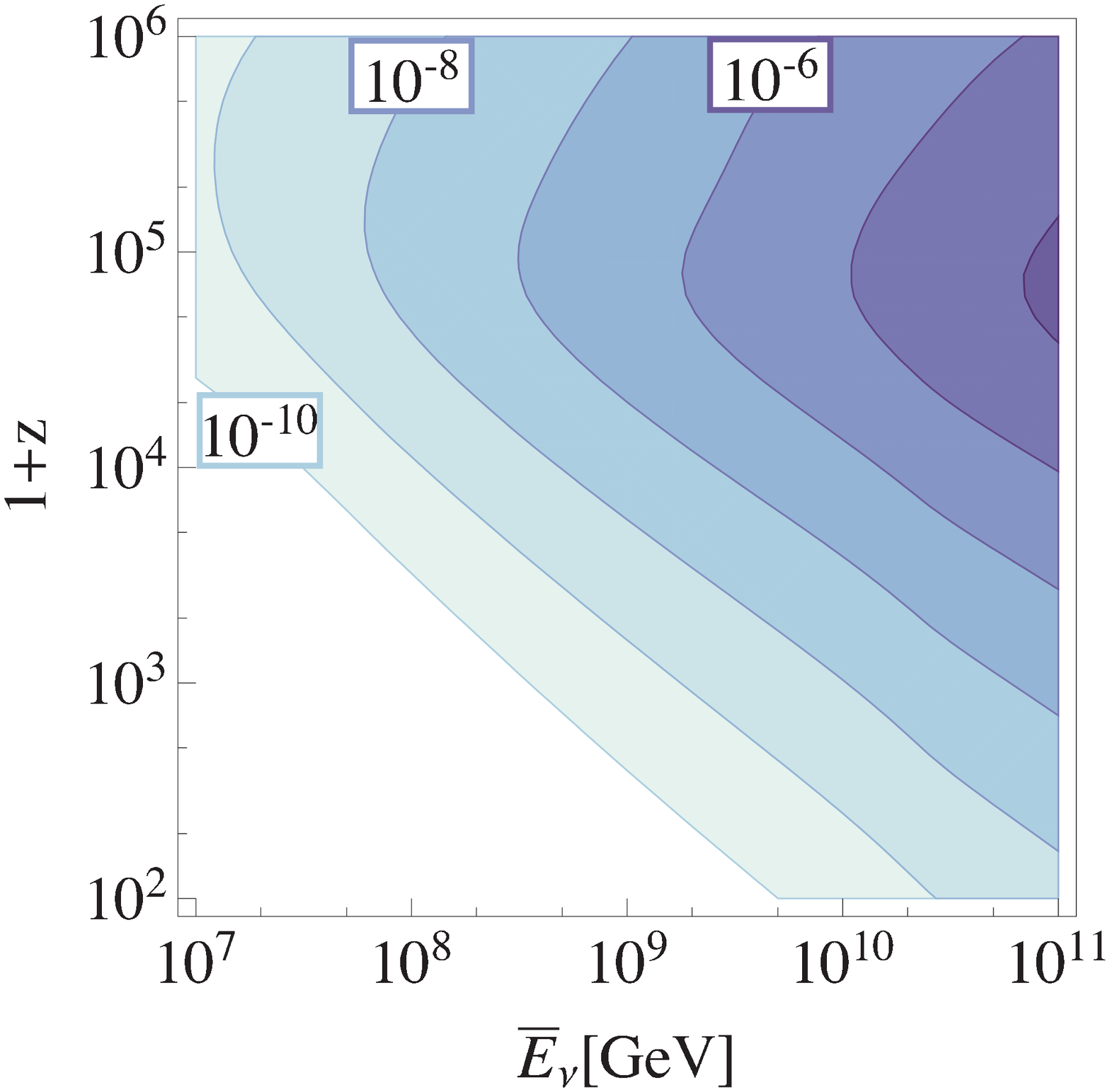}
    \caption{\small Contours of constant $\mu$ on $\bar{E}_\nu$ vs.\
      $1+z_*$ plane.  The contours are $\mu=10^{-10}$, $10^{-9}$,
      $\cdots$, and $10^{-4}$, from left to right.  (The numbers in the
      figure give the value of $\mu$.)  Here, we take $Y_{X} =
      10^{-22}$.}
    \label{fig:mu-param}
  \end{center}
\end{figure}

We calculate the $y$ and $\mu$ parameters as functions of
$\bar{E}_{\nu}$, $z_{*}$, and $Y_{X}$.  In Figs.\ \ref{fig:y-param}
and \ref{fig:mu-param}, we show the contours of constant $y$ and $\mu$
for $Y_{X}=10^{-22}$ on $\bar{E}_\nu$ vs.\ $1+z_*$ plane.  

For $E \lsim 10^{7}\ \text{GeV}$, both the $y$-type and $\mu$-type
distortions are almost negligible. This is because neutrinos are very
transparent for $E \lsim 10^{7}\ \text{GeV}$ and $1+z_{*}\lsim
10^{6}$. The $y$-type and $\mu$-type distortions become important only
when a significant amount of secondary photons and charged particles
are produced by the neutrino scattering.

As the energy of the neutrino becomes larger, $y$ and/or $\mu$ may
become sizable.  For $1+z_{*}\lsim 10^{4}$, $y$ is larger for larger
$1+z_{*}$ or $\bar{E}_{\nu}$. This is because the neutrino scattering
is more efficient with larger $1+z_{*}$ or $\bar{E}_{\nu}$. For
$1+z_{*}\gsim 5 \times 10^{4}$, $y$ rapidly decreases as $1+z_{*}$
increases. This is mainly due to the fact that, for $1+z_{*} \gsim
5\times 10^{4}$, a large fraction of $X$ decays before $z=z_{K}$.  On
the contrary, the $\mu$-type distortion is important only when
$1+z_{*}\gsim 5\times10^{4}$.  The reason is that a significant amount
of $X$ must decay when $z>z_{K}$ to realize sizable $\mu$. One can
also see that, at $1+z_{*}\gsim10^{5}$, $\mu$ becomes suppressed with
the increase of $z_{*}$. For $E\gsim 10^{7}\ \text{GeV}$ and
$1+z_{*}\gsim10^{5}$, $\mu$ can be approximately estimated as the
ratio of the energy density of $X$ to that of radiation at
$z=z_{*}$. With $\bar{E}_{\nu}$ and $Y_{X}$ being fixed, the energy
density of $X$ is proportional to $(1+z)^{3}$, while that of radiation
energy density scales as $(1+z)^{4}$. Therefore the ratio at $z=z_{*}$
is proportional to $(1+z_{*})^{-1}$, resulting in the fact that $\mu$
is also proportional to $(1+z_{*})^{-1}$ as far as $1+z_*\lesssim
  10^6$.

\subsection{Effects on the BBN}
\label{sec:BBN}

Finally, we consider the effects of the high-energy neutrino injection
on the BBN.  Due to the injection of hadrons and electromagnetic
particles as a consequence of the scattering processes of high-energy
neutrinos, hadronic and electromagnetic showers are induced.  Energetic
particles in the shower scatter off the light elements generated by
the BBN reactions, which results in the change of light-element
abundances.  Using the fact that the standard BBN scenario predicts
light-element abundances which are more-or-less consistent with
observations, scenarios with too much injections of hadrons and
electromagnetic particles are excluded.

In the following, we consider the case where $z_*$ is smaller than
$\sim 10^6$, for which photodissociation processes become important.
In particular, the overproduction of $^3$He due to the dissociation of
$^4$He provides the most stringent constraint; in our analysis, we
adopt the following bound \cite{Kawasaki:2004yh,
  Kawasaki:2004qu}:\footnote
{For the BBN constraints on the neutrino injection with smaller
    $\bar{E}_\nu$ than the present case, see \cite{Kawasaki:1994bs,
      Kanzaki:2007pd}.}
\begin{align}
  E_{\rm vis} Y_X < 2\times 10^{-14}\ {\rm GeV},
  \label{BBN}
\end{align}
where $E_{\rm vis}$ is the total energy injection in the form of
electromagnetic particles due to the decay of one $X$.  We have
estimated $E_{\rm vis}$ by using the energy injection rate at the
cosmic time being $\tau_X$.

\section{Constraints on Neutrino Emission}
\label{sec:constraints}
\setcounter{equation}{0}

Now, we are at the position to derive constraints on the early-decay
scenario with neutrino emission.  Here, we derive upper bounds on the
yield variable $Y_{X}$ by using the constraints from observations.  In
the present scenario, we take account of the following
constraints: (i) observational bounds on the high-energy
neutrino flux, (ii) bounds from the CMB spectral
distortions as we discussed in Sec.\ \ref{sec:distortion}, and
(iii) bounds from the BBN as we
discussed in Sec.\ \ref{sec:BBN}.

\subsection{Observational Constraints}
\label{sec:const}

We first consider the constraints from the neutrino flux.  To put
bounds on $Y_X$, we adopt the following upper bounds on the neutrino
flux:
\begin{itemize}
\item[(a)] For $E \leq 10^5\ \text{GeV}$, we take as the upper bound
  twice the atmospheric neutrino flux given in \cite{Honda:2006qj} and
  \cite{Abbasi:2011jx} (model 9 of Fig.~10 in \cite{Abbasi:2011jx}).
\item[(b)] For $10^5\ \text{GeV} < E \leq 10^6\ \text{GeV}$, we take 
$E^{2}\Phi_{\nu}(E) = 3.0 \times 10^{-8} (E/100\text{TeV})^{-0.3}$ 
  $\text{GeV}\text{cm}^{-2}\text{s}^{-1}\text{sr}^{-1}$ as the upper bound.
  This is twice the best-fit value of the neutrino flux for this energy region
  given by the IceCube collaboration \cite{Aartsen:2014gkd}.
  \item[(c)] For $10^6\ \text{GeV} < E \leq 10^{10}\ \text{GeV}$, we use
  the upper bound on the flux in this energy region given by the
  IceCube collaboration \cite{Aartsen:2013dsm}.
\end{itemize}

As we discussed in Sec.\ \ref{sec:distortion}, we may also obtain the
bound on $Y_X$ from the CMB spectrum distortion.  Currently, the
COBE/FIRAS experiment \cite{Fixsen:1993rd, Fixsen:1996nj} gives the
most stringent upper bounds on $y$ and $\mu$, which are
\begin{align}
|y| \leq 1.5 \times 10^{-5},
\end{align}
and
\begin{align}
|\mu| \leq 9\times10^{-5}.
\end{align}
We use these values to derive constraints on the yield variable
$Y_{X}$.

In addition, we consider the bound from the BBN.  The discussion below
takes account of the constraint given in Eq.\ \eqref{BBN}.

\subsection{Upper Bounds on $Y_X$}
\label{sec:results}

\begin{figure}[t]
  \begin{center}
    \includegraphics[width = 9cm]{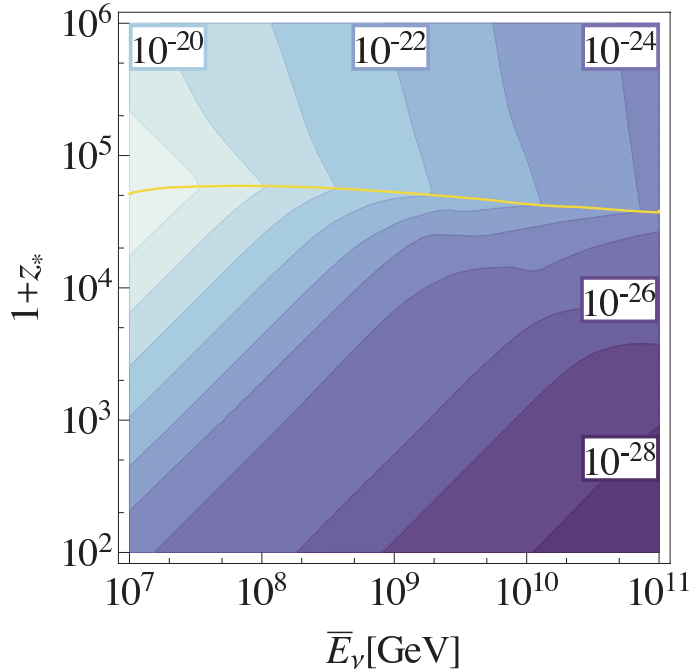}
  \end{center}
  \caption{\small The upper bound on $Y_{X}$ allowed by the current
    observations as a function of the neutrino energy $\bar{E}_{\nu}$
    at the emission and the typical redshift $z_*$ at the decay.  The
    contours are the upper bound on $Y_{X}$ equal to $10^{-16}$,
    $10^{-17}$, $\cdots$, and $10^{-28}$, from left to right.  (The
    numbers in the figure give the value of the upper bound.)  Below
    (above) the yellow line, the constraint from the high-energy
    neutrino flux (BBN) is stronger.}
    \label{fig:yield}
\end{figure}

Taking accounts of the observational bounds discussed in the previous
subsection, we derive the upper bound on the yield variable $Y_{X}$ as
a function of $\bar{E}_{\nu}$ and $z_{*}$.  In Fig.\ \ref{fig:yield},
we plot the upper bound on $Y_X$.  The bound comes from the
  neutrino flux (BBN) below (above) the yellow line; we find that the
  bound from the CMB distortion is currently weaker than that from
  BBN.

One can see that, for $1 + z_* \lsim 10^{4}$, the upper bound on
$Y_{X}$ depends only on the combination of $\bar{E}_{\nu}/(1+z_{*})$.
This is because the neutrino scattering processes are unimportant for
neutrinos produced at $1 + z \lsim 10^{4}$.  Then, the neutrino
flux is dominated by $\Phi_{\nu}^{(\text{prim})}$ given in Eq.\
\eqref{eq:flux}, which is sensitive to the combination of
$\bar{E}_{\nu}/(1+z_{*})$.  In such a region, we also note that the
observational constraints on the neutrino flux (a), (b), and (c) give
the most stringent bound on $Y_X$ for $\bar{E}_{\nu}/(1+z_{*})\lsim
10^{5}\ \text{GeV}$, $10^{5}\
\text{GeV}\lsim\bar{E}_{\nu}/(1+z_{*})\lsim 10^{6}\ \text{GeV}$,
and $\bar{E}_{\nu}/(1+z_{*})\gtrsim 10^{6}\ \text{GeV}$, respectively.
This can be understood from the fact that the present neutrino flux
has a peak at $E\sim \bar{E}_{\nu}/(1+z_{*})$ if the effects of the
neutrino scattering are negligible, as one can see from Fig.\
\ref{fig:present_flux}.

One can also see that, for $\bar{E}_{\nu} \gsim 10^{9}\ \text{GeV}$,
the constraint on $Y_{X}$ becomes weaker at around $1 + z_* \sim
10^{4}$. This is because the neutrino scattering processes are
effective in this region, resulting in the suppression of the
high-energy neutrino flux.  In addition, the constraints from 
the CMB spectral distortions and the BBN are not so stringent in this region.

We also consider the prospects of testing the present scenario, paying
particular attention to the possible improvement in the determination
of the $y$ and $\mu$ parameters in the future. 
For example, the PIXIE
experiment \cite{Kogut:2011xw} will offer much better sensitivity to
the CMB spectral distortions; $5\sigma$ detection is expected when
\begin{align}
  |y^{\rm (PIXIE)}| = 1\times10^{-8},
  \label{PIXIE-y}
\end{align}
or
\begin{align}
  |\mu^{\rm (PIXIE)}| = 5\times10^{-8}.
  \label{PIXIE-mu}
\end{align}

\begin{figure}[t]
  \begin{center}
    \includegraphics[width = 9cm]{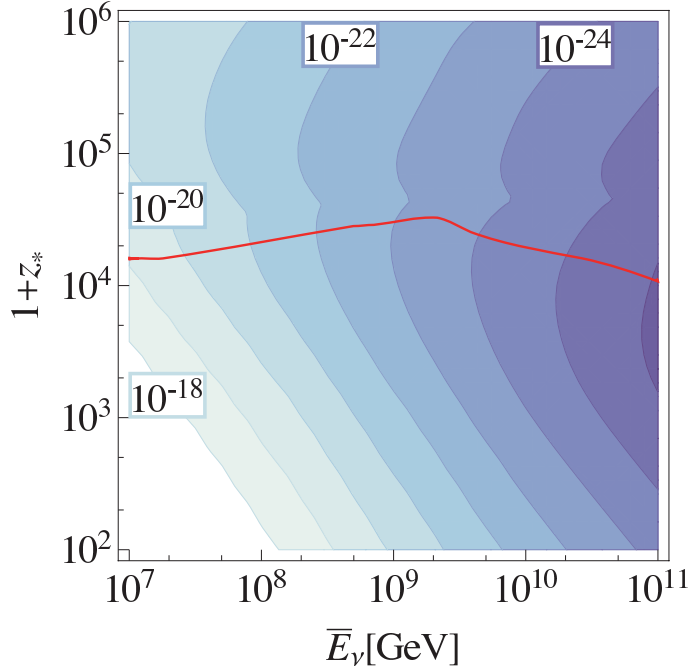}
  \end{center}
  \caption{\small The lower limit on $Y_{X}$ which can be detected by
    the PIXIE experiment at $5\sigma$ significance.  The contours are
    the lower bound on $Y_{X}$ equal to $10^{-16}$, $10^{-17}$,
    $\cdots$, and $10^{-28}$, from left to right.  (The numbers in the
    figure give the value of the lower bound for the detection.)
    Below the red line, the values shown in the figure are already
    excluded by the current observations of the cosmic-ray neutrino
    flux.}
  \label{fig:PIXIE}
\end{figure}

In Fig.\ \ref{fig:PIXIE}, we show the lower limit on $Y_{X}$ for the
$5\sigma$ detection with the expected PIXIE sensitivity.  The shape of
the contours reflects the dependences of $y$ and $\mu$ on
$(\bar{E}_{\nu}, z_{*})$ shown in Figs.\ \ref{fig:y-param} and
\ref{fig:mu-param}. There are kink-like structures at $1+z_{*}\sim
5\times10^{4}$. This is due to the fact that the injected energy is
equally distributed into $y$ and $\mu$ in our approximation for such a
redshift; the lowest value of $Y_{X}$ for the detection becomes
slightly weaker.  Then, above and below the kink, the value of $Y_X$
shown in Fig.\ \ref{fig:PIXIE} is obtained from the consideration of
$\mu$- and $y$-type distortions, respectively.

Even with the expected sensitivity of the PIXIE experiment, the
current bound on the neutrino flux provides better sensitivity to the
present scenario if $1+z_*\lsim 2 \times10^4$.  In Fig.\
\ref{fig:PIXIE}, the boundary of such a region is indicated by the red
line; in the region below the red line, the values of $Y_{X}$ shown in
the figure are already excluded by the current bounds on the neutrino
flux.  In other words, for $1+z_{*} \gsim 2 \times 10^{4}$ where the
neutrino scattering is efficient, future observations of the CMB
spectral distortions can test the parameter space which is not
explored by the current data.

We also note here that the PRISM experiment may provide another
accurate probe of the $y$- and $\mu$-parameters.  In
\cite{Andre:2013nfa}, it is claimed that the sensitivity of the PRISM
experiment can be as good as $\Delta\rho_{\rm rad}/\rho_{\rm rad}\sim
O(10^{-9})$, where $\Delta\rho_{\rm rad}$ is the total amount of the
energy release from decaying particles, which may correspond to a
better sensitivity than the PIXIE experiment.  With detection
sensitivities other than Eqs.\ \eqref{PIXIE-y} and \eqref{PIXIE-mu},
the value of $Y_X$ required for the detection of the signal can be
obtained by rescaling the values shown in Fig.\ \ref{fig:PIXIE}.

So far, we have assumed that $X$ dominantly decays into neutrinos.  If
electromagnetic particles are efficiently emitted by the decay of $X$,
however, we should also consider constraints from these decay
products.  We briefly comment on such a case although it is beyond the
scope of our study.  When $1+z_{*} \gsim 2\times10^{3}$, the
electromagnetic particles produced by $X$ contribute to the $\mu$-type
and $y$-type distortions or the dissociation processes of the light
elements after causing the electromagnetic cascade discussed in
Sec. \ref{sec:evol}.  For $\tau(z; E) \gsim 1$, the neutrino
scattering processes are so efficient that the neutrinos produced by
$X$ also induce the electromagnetic cascade.  As a result, if the same
amount of neutrinos and electromagnetic particles are produced by $X$,
the constraint on $Y_{X}$ due to the distortion of the CMB spectrum or
the light-element abundances is expected to be more-or-less unchanged
for such a parameter region.  For $\tau(z; E) \lsim 1$, however, the
constraint from electromagnetic particles due to the distortion of the
CMB spectrum or the light-element abundances is stronger than that
from neutrinos.  When $1+z_{*} \lsim 2\times10^{3}$, on the contrary,
electromagnetic particles produced by $X$ may change the ionization
history, which also gives the constraint on the injection of
electromagnetic particles \cite{Chen:2003gz, Zhang:2007zzh,
  Slatyer:2012yq}.

\section{Implication for Recent IceCube Result}
\label{sec:implication}
\setcounter{equation}{0}

\begin{figure}[t]
  \begin{center}
    \includegraphics[width = 0.95\textwidth]{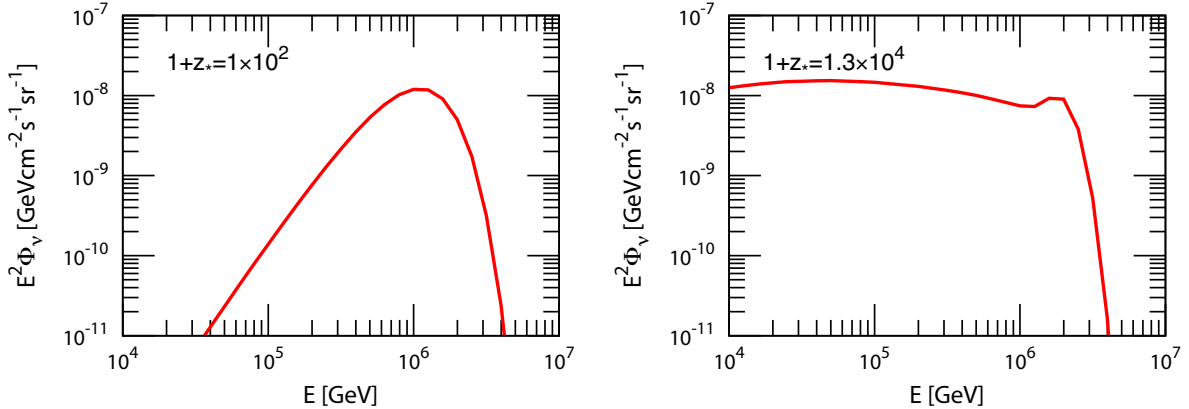}
    \caption{\small The present per-flavor neutrino fluxes for two
      different parameters. Left: $(\bar{E}_{\nu}, 1+z_{*}, Y_{X}) =
      (8\times10^{7}\ \text{GeV}, 10^{2}, 10^{-26})$ Right:
      $(\bar{E}_{\nu}, 1+z_{*}, Y_{X}) = (1.3\times10^{10}\
      \text{GeV}, 1.3\times10^{4}, 5\times10^{-26})$}
    \label{fig:IC}
  \end{center}
\end{figure}

In this section, we discuss the implications of the early-decay scenario
for the explanation of the origin of the high-energy cosmic-ray neutrinos 
observed by the IceCube collaboration.

Recently, the IceCube collaboration has published the results of their
three-year observation of high-energy neutrinos
\cite{Aartsen:2014gkd}. They detected three high-energy neutrino
events (nicknamed as Bert, Ernie, and Big Bird) with the deposited
energy of $1\ \text{PeV} \lsim E \lsim 2\ \text{PeV}$.  The number is
well above the expected background.  In addition to the PeV neutrino
events, the IceCube collaboration detected 34 events in the energy
region of $30\ \text{TeV} \lsim E \lsim 1\ \text{PeV}$, thus finding
37 events in total. Considering that the expected background is
$8.4\pm4.2$ from cosmic-ray muons and $6.6^{+5.9}_{-1.6}$ from
atmospheric neutrinos in this energy region \cite{Aartsen:2014gkd},
this gap suggests a new source of the energetic cosmic-ray neutrinos.
The IceCube collaboration claims that the per-flavor flux of
$E^{2}\Phi_{\nu}(E) = (0.95\pm 0.3) \times10^{-8}\
\text{GeV}\text{cm}^{-2}\text{s}^{-1}\text{sr}^{-1}$ in the energy
region of $60\ \text{TeV} < E < 3\ \text{PeV}$ is consistent with the
detected 37 events.\footnote{
For the detailed analysis of the standard-model interaction of neutrino 
with the detector, see also \cite{Chen:2013dza}.}
It is also claimed that, if the unbroken $E^{-2}$
power law spectrum is adopted, additional 3.1 events is expected above
$2\ {\rm PeV}$, while no event is observed in this energy region.  One
possibility is that the neutrino spectrum obeys $\sim E^{-2}$ power
law with the cutoff at the energy slightly above $\sim {\rm PeV}$
\cite{Aartsen:2014gkd}.

Because the origin of the high-energy cosmic-ray neutrino flux is yet
unknown, we pursue the possibility that the decay of an exotic
particle is responsible for it.  We will see that the three PeV
neutrino events at IceCube can be well explained in the present
scenario.  In addition, we will also see that $E^{-2}$ power law with
the cutoff at a few PeV may be realized, since the neutrino flux at
the energy higher than the position of the peak is exponentially
suppressed.

In Fig.\ \ref{fig:IC}, we show the present neutrino flux for two
sample points, which are given by
\begin{itemize}
\item Sample point 1 (left panel): $(\bar{E}_{\nu}, 1+z_{*}, Y_{X}) =
  (8\times10^{7}\ \text{GeV}, 10^{2}, 10^{-26})$,
\item Sample point 2 (right panel): $(\bar{E}_{\nu}, 1+z_{*}, Y_{X}) =
  (1.3\times10^{10}\ \text{GeV}, 1.3\times10^{4}, 5\times10^{-26})$.
\end{itemize}
The neutrinos produced by $X$ are very transparent for the case of the
sample point 1.  On the contrary, for the sample point 2, a sizable
amount of the initial neutrinos produced by $X$ is scattered and the
secondary neutrinos also contribute to the present neutrino flux. In
both sample points, the flux is $E^{2}\Phi_{\nu}(E) \sim 10^{-8}\
\text{GeV}\text{cm}^{-2}\text{s}^{-1}\text{sr}^{-1}$ at around PeV, so
they may explain the IceCube PeV events.

In the right panel in Fig.\ \ref{fig:IC}, one can also see that the
energy dependence of the flux is close to $E^{-2}$ for $E \lsim 2\
{\rm PeV}$. Therefore, with the parameters of our choice, there is a
possibility to explain all the IceCube events in the energy region of
$30\ \text{TeV} \lsim E \lsim 2\ \text{PeV}$ in the present scenario.
It should be, however, noted that the optical depth $\tau$ is very
sensitive to $\bar{E}_{\nu}$ and $z_{*}$ in the parameter region near
the sample point 2.  Therefore, the shape of the present neutrino flux
strongly depends on these parameters.  Requiring that the flux obeys a
power law of $E^{-2} - E^{-2.3}$ in the energy region of $60\
\text{TeV} < E < 3\ \text{PeV}$, for example, $z_*$ should be tuned
with the accuracy of $O(10\%)$ assuming that $\bar{E}_{\nu}\sim
10^{10}\ \text{GeV}$.  We emphasize here that, in order to realize the
$\sim E^{-2}$ power law, the neutrino scattering processes should
become efficient, which predicts sizable $y$ (or $\mu$).  In the
sample point 2, for example, $y = 3.0\times10^{-9}$ (and $\mu =
6.4\times10^{-10}$).  Such a value of the $y$ parameter is close to or
within the reach of the expected sensitivity of the PIXIE and the
PRISM experiments.  Hence if the IceCube events in the energy region
of $30\ \text{TeV} \lsim E \lsim 2\ \text{PeV}$ are explained in the
present scenario, the future experiments may have a chance to see the
CMB spectral distortion.

Here, we also point out that in order to explain the IceCube events in
the present scenario, $\bar{E}_\nu$ cannot be arbitrary large.  We
found that, for $\bar{E}_{\nu} \gsim 5\times10^{10}\ \text{GeV}$, the
flux of $E^{2}\Phi_{\nu}(E)|_{E=1\ \text{PeV}} \sim 10^{-8}\
\text{GeV}\text{cm}^{-2}\text{s}^{-1}\text{sr}^{-1}$ cannot be
obtained without conflicting the current observational constraints. 
Thus the mass of $X$ responsible for the
IceCube events should be smaller than $\sim 10^{11}\ \text{GeV}$.

Finally, we comment on the angular dependence of the neutrino
flux.  In the early-decay scenario, the neutrino flux is isotropic, which
is consistent with the IceCube result.  On the contrary, it has also
been discussed that the decay of dark matter may explain the IceCube
events \cite{Feldstein:2013kka,Esmaili:2013gha,
Bai:2013nga,Higaki:2014dwa,Bhattacharya:2014vwa,Zavala:2014dla,Chen:2014lla}.
In such a scenario, the Galactic contribution
dominates and large fraction of the energetic neutrinos is expected to
come from the direction of the Galactic center.  Denoting the angle
between the Galactic center and the direction of the neutrino as
$\theta$, the flux from $\theta < \pi/2$ is roughly twice as large as
that from $\theta > \pi/2$ at the peak of the flux \cite{Ema:2013nda}.
Future observations on the angular dependence may help to distinguish
the scenarios with $\tau_{X} \ll t_{0}$ and the ones with $\tau_{X} \gg t_{0}$.

\section{Conclusions and Discussion}
\label{sec:summary}
\setcounter{equation}{0}

In this paper, we have studied the cosmological implications of
high-energy neutrino injection from the decay of a massive particle
$X$.  When considering high-energy neutrinos in the early universe,
the scattering processes with background neutrinos are important. We
have numerically followed the evolution of the high-energy neutrino
flux including such neutrino scattering effects, and calculate the
present neutrino flux.  Importantly, even only via the weak
interaction, energetic neutrinos with $E\sim 10^8-10^{10}\ {\rm GeV}$
can effectively scatter off background neutrinos at $1+z\gsim
10^5-10^3$.  Such scattering processes affect the shape of the
cosmic-ray neutrino spectrum, as well as produce CMB distortions by
the emission of photons and charged particles.

We have derived the upper bounds on the yield variable $Y_X$ as a
function of the energy $\bar{E}_\nu$ of the neutrino emitted from the $X$ decay 
and $z_*\equiv z(\tau_X)$, using observational bounds on the high-energy
cosmic-ray neutrino flux and the CMB spectral distortions.  We
have seen that the former gives more stringent bound for the case
where the neutrino emission occurs when $1+z\lsim 10^4-10^5$.  In
particular, for $1+z_*\lsim 10^4$, the neutrino
scattering processes are irrelevant, and the upper bound on $Y_X$
depends only on the combination of $\bar{E}_{\nu}/(1+z_{*})$; in such
a case, we found that the bounds are
\begin{itemize}
\item $Y_X\lsim 9\times 10^{-23}$ for
  $\bar{E}_{\nu}/(1+z_{*})=10^4\ {\rm GeV}$,
\item $Y_X\lsim 2\times 10^{-25}$ for $\bar{E}_{\nu}/(1+z_{*})=10^5\ {\rm
    GeV}$,
\item $Y_X\lsim 1\times 10^{-26}$ for $\bar{E}_{\nu}/(1+z_{*})=10^6\ {\rm
    GeV}$,
\item $Y_X\lsim 8\times 10^{-28}$ for $\bar{E}_{\nu}/(1+z_{*})=10^7\ {\rm
    GeV}$.
\end{itemize}
On the other hand, for $1 + z_* \gsim 10^5$, BBN
gives a stronger bound.  
With the current accuracy, the CMB bound is less stringent than the
BBN bound.  However, with the sensitivity of the future experiments,
PIXIE and PRISM, for example, the upper bound from CMB observation is
expected to be improved by about three orders of magnitude, which will
give stronger bound than the BBN.

We have also considered the possibility that the PeV neutrino events
recently observed by IceCube originate from the decay of $X$.  We have
seen that the three PeV neutrino events (Bert, Ernie, and Big Bird)
can be well explained within this scenario.  In addition, we have seen
that, when the decay of $X$ occurs at $1+z_*\sim 10^{4}$ and the
initial energy of neutrino produced by the decay of $X$ is $\sim
10^{10}\ \text{GeV}$, we have a possibility to realize an $E^{-2}$
power law neutrino spectrum with a cutoff at $\sim$ PeV, which is
suggested by the IceCube results; for such a scenario, $z_*$ should be
tuned with the accuracy of $O(10\ \%)$.  In addition, future
observation of the CMB may be able to detect the distortion of the
CMB spectrum caused by the decay of $X$.

We emphasize here that the mass of $X$ responsible for the PeV
neutrino events can be as large as $O(10^{10}\ \text{GeV})$, if
$1+z_*\sim 10^4-10^5$ (which corresponds to $\tau_X\sim10^{11}-10^9\
{\rm sec}$).  In other words, the IceCube experiment can probe the
physics at the energy scale much higher than PeV.  One of the examples
of the new physics containing the candidate of the massive particle
$X$ is the model with Peccei-Quinn symmetry \cite{Peccei:1977hh,
  Peccei:1977ur} because the natural scale of the Peccei-Quinn
symmetry breaking is $O(10^{9-10}\ \text{GeV})$.  Another possibility
can be a messenger sector in gauge-mediated supersymmetry breaking
model \cite{Dine:1993yw, Dine:1994vc, Dine:1995ag}.  More discussion
about particle-physics models with the candidates of the massive
particle $X$ is found, for example, in \cite{Ema:2013nda,
  Feldstein:2013kka, Higaki:2014dwa}. Therefore, the future IceCube
experiment can shed light not only on astrophysical sources of
cosmic-ray neutrinos, but also on high-energy particle-physics.

\vspace{1em}
\noindent {\it Acknowledgements}: One of the authors (T.M.) is
grateful to the Mainz Institute for Theoretical Physics (MITP) for its
hospitality and its partial support during the completion of this
work. 
R.J. is supported by the JSPS fellowship (No.~25-8360).  
T.M. is supported by the JSPS KAKENHI (No.~26400239 and No.~60322997).
The work of Y.E. and R.J. is also supported by the Program for
Leading Graduate Schools, MEXT, Japan.



\end{document}